\documentclass[11pt,twoside]{article}

\usepackage{asp2006}
\usepackage{epsf}
\usepackage{psfig}
\usepackage{lscape}

\begin{document}
\title{Precise Photometry and Spectroscopy of Transits}   
\author{Joshua N.\ Winn}   
\affil{Massachusetts Institute of Technology,
77 Massachusetts Avenue, Cambridge, MA 02139}    

\begin{abstract} 
  A planetary transit produces both a photometric signal and a
  spectroscopic signal. Precise observations of the transit light
  curve reveal the planetary radius and allow a search for timing
  anomalies caused by satellites or additional planets. Precise
  measurements of the stellar Doppler shift throughout a transit (the
  Rossiter-McLaughlin effect) place a lower bound on the stellar
  obliquity, which may be indicative of the planet's migration
  history. I review recent results of the Transit Light Curve project,
  and of a parallel effort to measure the Rossiter effect for many of
  the known transiting planets.
\end{abstract}


\section{Introduction}   

I have great admiration for the people who discover transiting
planets. Identifying the candidate transit signals from among a
hundred thousand light curves, and flushing out the numerous
astrophysical false positives, are impressive feats. This article,
however, is not about transit discovery, but rather about the next
step: performing high-precision photometry and spectroscopy of
exoplanetary transits. The goal of this step is to determine the
planetary and stellar properties well enough to allow for meaningful
comparisons with the familiar properties of the Solar system, and to
inform our theories of planet formation.

The most immediate result of transit photometry is a measurement of
the planetary radius. In combination with the planetary mass, which
can be inferred from the Doppler orbit of the star, these data give
the first clues about the composition, interior structure, and
atmospheric energy balance of the planet. An accurate radius is also
needed to interpret the results of other observations, such as the
detection of thermal emission or reflected light based on
secondary-transit photometry. The timings of the transits can be used
to refine the measurement of the orbital period and search for
additional bodies in the system. In \S~2, I describe the Transit Light
Curve (TLC) project, an effort to gather high-precision photometry
during exoplanetary transits.

The most prominent spectroscopic signal during a transit is the
Rossiter-McLaughlin (RM) effect. This effect is an anomalous Doppler
shift that arises from stellar rotation. Measuring this effect allows
one to assess the alignment between the planetary orbital axis and the
stellar spin axis, a fundamental system property that provides clues
about the process of planet migration. I describe some recent
measurements of the RM effect in \S~3.

\begin{figure}[!h]
\plotone{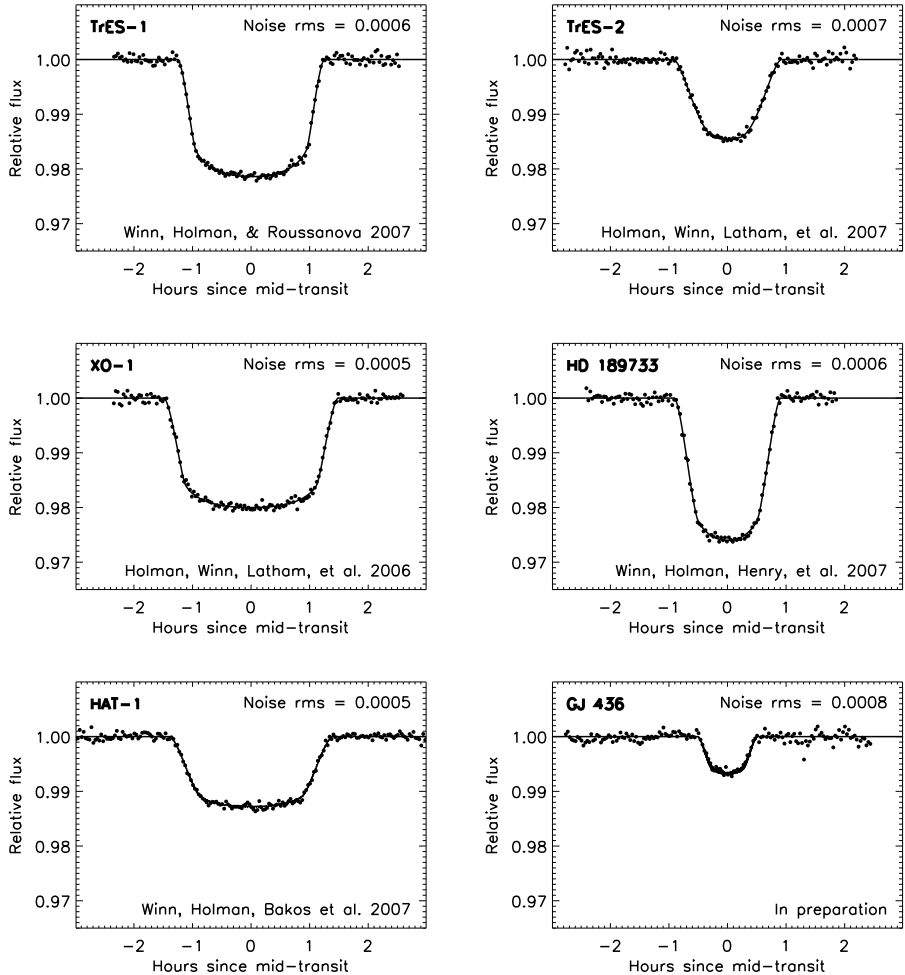}
\caption{{\bf The Transit Light Curve project.} Each panel shows
   time-binned photometry from a campaign (2-5 separate transits) for
   a particular planet.  Most of the data are $z$ band observations
   with the FLWO 1.2m telescope and Keplercam detector.}
\end{figure}

\section{The Transit Light Curve project}   

The most spectacular transit light curves have come from space-based
observations with the {\it Hubble Space Telescope}\, (Brown et
al.~2001, Pont et al.~2007) and the {\it Spitzer Space Telescope}\,
(Knutson et al.~2007). However, it is important to remember that the
extreme precision with which the relative flux can be recorded
($\sim$$10^4$~per 30~s sample) does not guarantee an extremely precise
value for the planetary radius. Rather, the transit depth $\Delta F
\approx (R_p/R_\star)^2$ is known precisely, but $R_p$ is known only
as well as $R_\star$. The stellar radius must be estimated using other
observable properties of the star, such as its parallax, luminosity,
angular diameter, or spectrum.

Actually the situation is slightly more complex. In addition to the
transit depth, precise photometry provides a few other observables,
among which are the photometric period $P$, the total transit duration
$t_T$ (from first to fourth contact), and the full transit duration
$t_F$ (from second to third contact). Kepler's third law can be used
to write an accurate expression for the mean stellar density in terms
of observable quantities (Seager \& Mallen-Ornelas 2003):
\begin{equation}
\rho_\star = \frac{32}{G\pi}~P~
             \frac{ \Delta F^{3/4} }{ (t_T^2-t_F^2)^{3/2} }
\end{equation}
This approximation assumes $M_p \ll M_\star$, $R_p \ll R_\star \ll a$,
and a circular orbit, but it is straightforward to generalize this
expression. Thus, precise photometry reveals the mean stellar density.

This is helpful because it means that the light curve itself helps to
estimate the stellar radius. A traditional method for estimating the
stellar radius is to compare the spectroscopic values of $T_{\rm
  eff}$, $\log g$, and metallicity with the output of theoretical
models of stellar evolution. With excellent transit photometry it is
advantageous to use $\rho_\star$ instead of $\log g$ (see, e.g.,
Sozzetti et al.~2007, Holman et al.~2007). A related point is that in
some regions of the HR diagram, the stellar {\it mass}\, is
constrained at least as well as the stellar radius. At fixed stellar
density---that is, with a perfect transit light curve---the error in
$R_\star$ (and hence $R_p$) varies only as the cube root of
$M_\star$. In such cases the systematic error in $R_p$ would be
reduced by a factor of three relative to a situation in which only the
transit depth (and not the durations) were known.

Another quantity that can be written purely in terms of observables is
the planetary surface gravity (Southworth et al.~2007). In this sense,
what a transit observer {\it really}\, measures is the stellar mean
density and the planetary surface gravity. Transforming these
variables into $R_p$ and $M_p$ requires external information.

In 2006, Matt Holman and I wondered whether it would be possible to
reach the limiting precision in $R_p$ by combining the results of
repeated ground-based observations. We began the Transit Light Curve
(TLC) project, with the aim of building a library of transit
photometry of uniformly high quality, along with unified and rigorous
methods for parameter determination. Our workhorse is the Fred
L.~Whipple~1.2m telescope and Keplercam detector. We have also used
the 6.5m Magellan telescopes for the OGLE systems, which are both
southerly and faint. We have found that by employing these instruments
and some straightforward observing protocols, it is possible in some
cases to reach the limit in which the error in $R_p$ is dominated by
the uncertainties in the stellar properties rather than the
photometric uncertainties.

Among our observing protocols are the following. When possible we
center the field in such a way as to encompass at least 5 comparison
stars of similar brightness to the target star. We keep the image
registration as consistent as possible throughout the night. We
defocus, if needed, to broaden the point-spread function to a
consistent width of 5-7 pixels; this often has the salutary effect of
enhancing the duty cycle by lengthening the maxmimum exposure time.
We ensure that the flat-field calibration frame has enough counts to
correct for $\sim$0.1\% pixel-to-pixel sensitivity variations. We
observe not only the entire transit but also at least 1~hr prior to
ingress and 1~hr after egress. We use the reddest available optical
bandpass (often Sloan $z$), for two reasons: to minimize the effect of
differential extinction on the relative photometry, and to minimize
the effect of limb darkening on the transit light curve.

Figure~1 is a gallery of some of our composite light curves, along
with the standard deviation of the noise and citations to published
results. Figure~2 shows the favorable noise characteristics that we
have found for a subset of our target systems. Naturally, the best
results are obtained when the field of view offers numerous good
comparison stars and when the sky conditions are pristine. (Even
relative photometry benefits from ``photometric'' nights.) A
photometric precision of 0.15\% with 30-second sampling is typical.

\begin{figure}[h]
\plotone{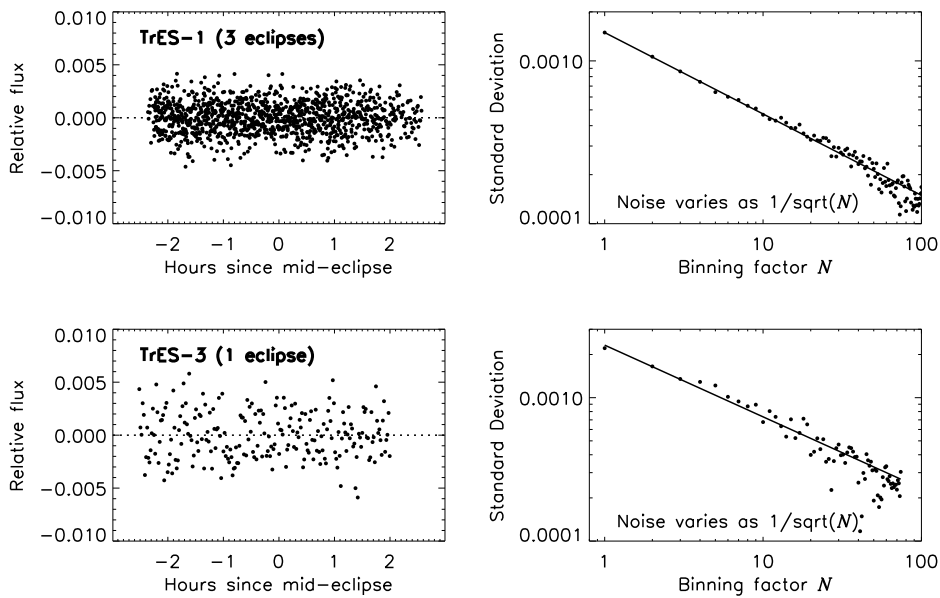}
\caption{Photometry of TrES-1 and TrES-3, after subtracting the
   best-fitting model. The left panels show the residuals as a
   function of time. The right panels show the standard deviation of
   the binned residuals, as a function of the bin size.  For these
   systems the noise varies as $N^{-1/2}$ as one would expect for
   uncorrelated Gaussian noise. The data sets exhibiting these
   favorable characteristics generally come from observations on
   photometric nights of systems for which 5 or more good comparison
   stars are available.}
\end{figure}

We have written a comprehensive analysis code that simultaneously fits
multiple transits observed in different bandpasses, secondary transits
(when such data are available), and radial velocities (including the
Rossiter-McLaughlin effect; see \S~3).  We use a Markov Chain Monte
Carlo algorithm to estimate the joint {\it a posteriori}\, probability
distribution for all relevant parameters.  Going forward, we intend to
make some potentially important improvements to the code, including a
more sophisticated treatment of ``red noise'' employing a model of the
covariance matrix (at the moment we use a crude approxmation), and a
principal-component decomposition of the limb darkening function.

Over the longer term, we hope to detect (or rule out) additional
planets and satellites of the transiting planet. These may manifest
themselves through gradual changes in the transit duration caused by
orbital precession (Miralda-Escud\'e~2002, Heyl \& Gladman~2007) and
by short-term variations in the midtransit times due to
multi-Keplerian motions or mutual gravitational interactions.  We are
gathering a large number of transit times with a precision of
0.25-1~min. Holman \& Murray (2005) and Agol et al.~(2005) have shown
that transit-timing variations of that order could be produced by
perturbers as small as the Earth, as long as they are in mean-motion
resonances with the transiting planet. Resonances might be a common
outcome of planet formation and migration, and searching for
transit-timing variations is one way to find out.

\section{The Rossiter-McLaughlin Effect}   

A striking pattern in the Solar system is its coplanarity. The orbital
axes of the 8 planets line up to within a few degrees, and the Sun's
spin axis differs by only 7 degrees from the Earth's orbital axis. The
observed coplanarity was the original inspiration for the theory that
the Sun and planets condensed from a single spinning disk. It would be
interesting to know whether this degree of alignment is typical of all
planetary systems. Exoplanets have provided enough surprises that
nothing should be taken for granted.

Indeed there are reasons to expect at least occasional misalignments.
Among the surprises to which I just alluded is that planetary orbital
eccentricities are often large; whatever mechanism perturbs orbital
eccentricities may also perturb inclinations. For close-in planets,
which are thought to have formed at large orbital distances and then
migrated inward, one may wonder whether the migration process
disturbed the original alignment. The various migration theories
differ on this point. Migration via tidal interactions with the
protoplanetary disk should not perturb the alignment and may even
drive the system toward closer alignment (Ward \& Hahn 2003). In
contrast, migration via planet-planet scattering would magnify any
initial misalignments (Chatterjee, Ford \& Rasio 2007), and migration
via Kozai cycles accompanied by tidal circularization results in a
broad distribution of final inclination angles (Fabrycky \& Tremaine
2007). Thus, measuring the alignment between the orbital axis and the
stellar spin axis is a means for testing migration theories and for
identifying planets that migrated in interesting ways.

Transit observations can be used to assess the spin-orbit alignment,
courtesy of a spectroscopic effect observed long ago by Rossiter
(1924) and McLaughlin (1924) in eclipsing binaries. During a transit,
the planet hides part of the rotating stellar disk, causing an
``anomalous'' Doppler shift. When the planet transits the approaching
(blueshifted) half of the star, the net starlight appears slightly
redshifted, and vice versa, as shown in Fig.~3. For the configuration
shown, in which the transit impact parameter is approximately 0.5, the
signal for a well-aligned planet is antisymmetric about the midtransit
time, whereas a strongly misaligned planet that spends all its time
covering the receding half of the star would produce only an anomalous
blueshift. The key parameter is $\lambda$, the angle between the sky
projections of the orbital axis and the stellar rotation axis.

\begin{figure}[!h]
\plotone{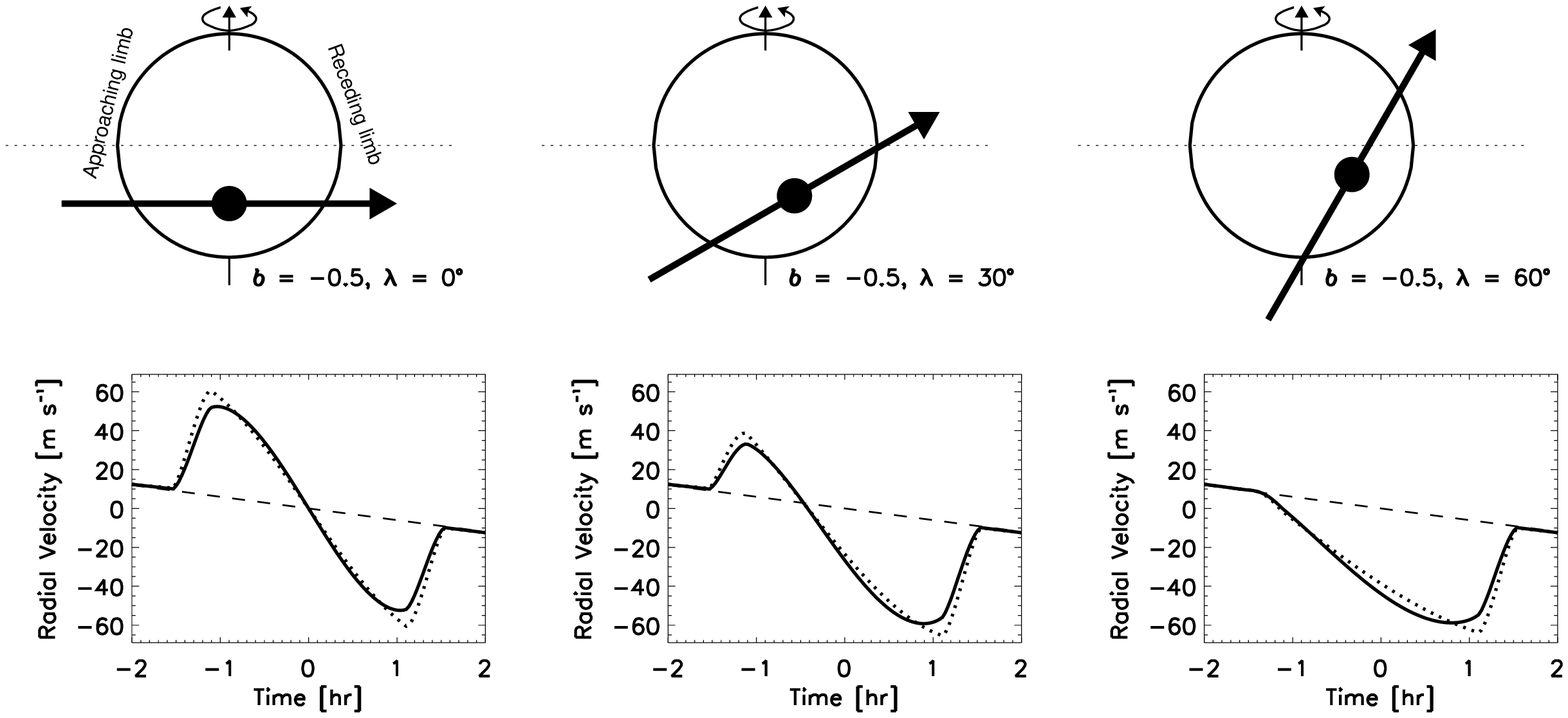}
\caption{ {\bf The Rossiter-McLaughlin effect.} Shown are three planet
  trajectories that produce identical light curves, but have different
  orientations relative to the stellar spin axis and hence produce
  different Rossiter-McLaughlin signals. In the bottom panels, the
  dotted lines show a model of the effect in the absence of limb
  darkening; the solid lines show a model that includes limb
  darkening. Adapted from Gaudi \& Winn (2007). }
\end{figure}

Measurements of the projected spin-orbit angle $\lambda$ have been
published for 5 systems (Queloz et al.~2000, Winn et al.~2005, 2006,
2007; Wolf et al.~2007; Narita et al.~2007). In all cases $\lambda$ is
consistent with zero at the 1-2$\sigma$ level, with accuracies ranging
from 1--30 degrees. This has led to the working hypothesis that the
migration of hot Jupiters generally preserves spin-orbit alignment.
Importantly, star-planet tidal interactions (which are responsible for
spin synchronization and orbital circularization of the planet) do not
confound the interpretation because the expected timescale for tidal
reorientation is generally $10^{10}$~yr or longer, as estimated using
the equilibrium-tide theory of Hut~(1981).

\begin{figure}[!h]
\plotone{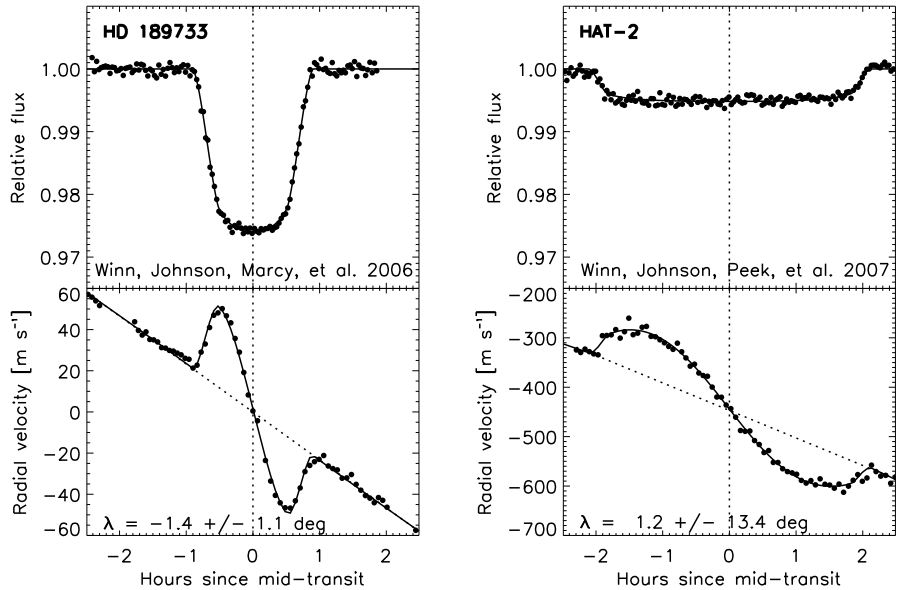}
\caption{ Photometric and spectroscopic observations of transits of
   HD~189733 (left) and HD~147506 (right; also known as HAT-P-2). The
   radial velocity data during the transit exhibit the
   Rossiter-McLaughlin effect.  The projected spin-orbit angle
   $\lambda$ was found to be small in both cases, through a
   simultaneous fit to the photometry and radial-velocity data. The
   precision is much lower for HD~147506 because the transit is nearly
   equatorial, which causes a strong degeneracy between $v\sin i$ and
   $\lambda$ (see Gaudi \& Winn 2007).}
\end{figure}

Observations of the Rossiter-McLaughlin effect provide only a lower
limit on the stellar obliquity, the angle between the orbital and
rotation axes. The RM waveform is sensitive the angle between the sky
projections of those axes, and the inclinations with respect to the
sky must be determined with other data. The orbital inclination is
often known to within $\approx$1~deg from the transit light curve, but
the stellar inclination is usually unknown. The single exception thus
far is HD~189733. That star is chromospherically active and exhibits
quasiperiodic flux variations (presumably due to star spots) from
which the rotation period can be measured (Henry \& Winn 2007). The
combination of a measured rotation period, stellar radius, and
projected rotation rate (which can be measured from either the
amplitude of the RM effect or from the observed spectral-line
broadening) places a constraint on the stellar inclination, which is
consistent with being edge-on.

More work remains to be done on confronting the specific predictions
of planet-planet scattering and Kozai cycles with the data.  The case
of HD~147506 (also known as HAT-P-2) is especially interesting because
the planet has a highly eccentric orbit ($e=0.5$), naturally raising
the possibility of planet-planet scattering or Kozai cycles, yet the
system is well-aligned (Winn et al.~2007). In this case, however, the
achievable accuracy in $\lambda$ was limited to about 14 degrees
because the transit is nearly equatorial. In such cases the Rossiter
waveform is always antisymmetric and the value of $\lambda$ is encoded
in the {\it amplitude}\, of the signal, which causes a strong
degeneracy with the value of the stellar projected rotation velocity
(see Gaudi \& Winn 2007).

\acknowledgements 
It is a pleasure to thank Matt Holman, my partner in the Transit Light
Curve project; and John Johnson, Scott Gaudi, Geoff Marcy, Norio
Narita, Yasushi Suto, and Ed Turner, for their important contributions
to the Rossiter program. I am very grateful to Fred Rasio, Alex
Wolszczan, and the other conference organizers for the opportunity to
share this work in a memorable setting on Santorini.


\end{document}